\documentclass [11pt]{amsart}
\usepackage{amscd}
\usepackage{amsxtra}
\usepackage{amsthm}
\usepackage{epsfig}
\usepackage{graphicx,color}
\usepackage{amssymb}
\usepackage[all]{xy}
\usepackage{mathbbol}
\usepackage[ansinew]{inputenc}
\usepackage[T1]{fontenc}
\newtheorem{theorem}{\sc Theorem}

\newtheorem{proposition}{\sc Proposition}[section]
\newtheorem{lemma}{\sc Lemma}[section]
\newtheorem{corollary}{\sc Corollary}[section]
\theoremstyle{remark}
\newtheorem{definition}{\sc Definition}[section]
\newtheorem{remark}{\it Remark}[section]
\newtheorem{example}{\it Example}[section]

\font\tmsb=msbm10 at12pt
\font\smsb=msbm7
\font\ssmsb=msbm5
\newfam\msbfam
\textfont\msbfam=\tmsb
\scriptfont\msbfam=\smsb
\scriptscriptfont\msbfam=\ssmsb
\def \1{\mathbb {1}}
\def \RM{\mathbb {R}}
\def \ZM{\mathbb{Z}}
\def \CM{\mathbb{C}}
\def \Cq{\mathbb{C}_{\hbar }}

\def \Kt {\mathcal{K}}
\def \HM {\mathbb{H}}

\def \Bt {\mathcal{B}}
\def \Ct {\mathcal{C}}

\def \Et {\mathcal{E}}
\def \Ht {\mathcal{H}}

\def \Hom {{\rm Hom}}
\def \Sp {{\rm Sp}}
\def \d{\partial}

\def\dt{\delta} 
\def\a{\alpha}
\def\b{\beta}
\def\e{\varepsilon}  
\def\g{\gamma}
\def\p{\varphi}

\def\l{\lambda}

\def\lb{\left\{}
\def\rb{\right\}}

\def \s{\sigma}
\def \t{\tilde}

\def \to{\longrightarrow} 

\def \< {{\langle }}
\def \> {{\rangle }}

\newcommand{\M}{{\mathcal M}}

\newcommand{\Ft}{{\mathcal F}}

\newcommand{\OM}{{\mathcal O}}

\newcommand{\Qt}{{\mathcal Q}}
\newcommand{\Lt}{{\mathcal L}}

\begin{document}
\title [Perturbative expansions in quantum mechanics] {
Perturbative expansions in \\ quantum mechanics}
\author[Mauricio D. Garay]{Mauricio D. Garay$^*$}
\thanks{$*$ partly supported by the Forschungsstipendium GA 786/1-1
of the Deutsche Forschungsgemeinschaft and by the
IH{\'E}S ($6^{th}$ European Framework program,
contract Nr. RITA-CT-2004-505493).}
\date{Original version February 2005, Revised May 2007}
\address{SISSA/ISAS, via Beirut 4, 34014 Trieste,
Italy.}
\email{garay@sissa.it}


\parindent=0cm


\begin{abstract}{We prove a $D=1$ analytic versal deformation theorem
for WKB expansions.
We define the spectrum of an operator in local analytic terms. We use
the Morse lemma to show that the perturbation series arising in a perturbed harmonic oscillator become
analytic after a formal Borel transform.}
\end{abstract}
\maketitle
\section*{Introduction}
In this paper, we study deformation theory in an analytic subalgebra $\Qt$
 of the universal Heisenberg algebra $\CM[[x,\hbar \d_x]]$. The algebra $\Qt$ 
consists of analytic micro-differential operators depending on a
 semi-classical parameter $\hbar$. We prove a Morse lemma in the
 $\Qt$-algebra, give further a versal deformation theorem and solve a
 conjecture formulated by Colin de Verdi{\`e}re (\cite{Colin}, Question 7).
 Next, we introduce a purely analytic definition of the spectrum
which coincides with perturbative computations. 
Then, using the quantum Morse lemma, we get that the perturbation
series for the spectrum of a perturbed harmonic oscillator
$H(q,p)=p^2+q^2+tg(t,q,p)$, such as the one computed by Heisenberg (\cite{Heisenberg}), are
{\em Borel analytic}.\\
By Borel analytic, we mean that these series can be
expanded in formal power series $ E(t,\hbar )=\sum_k \a_k(t)\hbar^k$
such that the corresponding {\em  formal Borel transforms}
$ \hat E(t,\hbar )=\sum_k 
\a_k(t)\hbar^k/k!$ are convergent for $|\hbar|$ sufficiently small,
i.e., $E(t,\hbar )$ is of Gevrey class $1$ or $2$ depending on
conventions.\\
Several examples of Borel analyticity are treated
in the literature among which the case of a perturbation given by
$g(t,q,p)=q^4$ (\cite{Pham_resurgence,Simon}).
For some special polynomial
perturbations, it is conjectured that the series are resurgent
(\cite{Pham_resurgence,Voros_exact,Zinn_Justin_nuclear}).\\
The results of this paper can be generalised, in higher dimensions, to quantum integrable systems
 (\cite{Arnold_70}).
The generalisation to Stein neighbourhoods of arbitrary compact subsets in $\CM^{2n}$ and the specialisation of the results to the real analytic case are
straightforward. 
\section{The quantum Morse lemma }

\subsection{The $\Qt$-algebra}
\label{SS::definitions}
Let $\widehat \Qt$ be the non-commutative
algebra consisting of formal power series in the variables
$a,a^\dag,\hbar$ which satisfy the commutation relations 
$$[a,a^\dag]=\hbar,\ [\hbar ,a]=0,\ [\hbar ,a^\dag]=0.$$
The operators $\frac{1}{\sqrt{\hbar}} a$ and $\frac{1}{\sqrt{\hbar}} a^\dag$
are the
annihilation and creation operators of a free bosonic theory. It is
important that unlike the classical approach, we do not divide these
operators by $\sqrt{\hbar}$.\\
We sometimes identify the algebra $\Qt$ with the algebra generated by 
$\hbar$ and the operators $p,q$ with
$$p=\frac{a^\dag+a}{\sqrt 2},\ q=\frac{a^\dag-a}{\sqrt 2i}$$
satisfying the commutation relation $[p,q]=-i\hbar$.\\
An element $f$ of the  $ \widehat \Qt $-algebra
can always be {\em ordered}, i.e., written as a formal sum
$f=\sum  \a_{mnk} (a^\dag)^ma^n\hbar^k$ with the $a^\dag$'s before the $a$'s.\\
The total symbol $s:\widehat \Qt \to \CM[[\hbar ,x,y]]$
is defined by replacing the variables $a^\dag,a$ with commuting variables $x,y$:
$$s(f)(\hbar,x,y)=\sum_{m,n,k \geq 0}  \a_{mnk} x^my^n\hbar^k.$$
The principal symbol $\s:\widehat \Qt \to \CM[[x,y]]$ is obtained by
restricting the total symbol to $\hbar=0$. 
We define the {\em Borel transform} $ B:\widehat \Qt \to \CM[[\hbar ,x,y]]$ by setting
$$ B(f):=\sum_{m,n,k \geq 0} \frac{ \a_{mnk}}{k!} x^my^n\hbar^k.$$
\begin{definition}[\cite{Pham_resurgence,Sjostrand_asterisque}]{\rm The {\em $\Qt$-algebra}
is the subalgebra of $\widehat \Qt$ consisting of power series having a convergent
Borel transform:
 $$\Qt=\lb f \in \widehat \Qt, Bf \in \CM \{\hbar ,x,y \} \rb.$$}
\end{definition}
Here the notation $ \CM \{\hbar ,x,y \}$ stands for the ring of absolutely
convergent power series.
The subalgebra of  $\Qt$ consisting of series which are independent of
$a^\dag$ and $a$ is denoted by $\Cq$:
$$\Cq=\{ \a \in \CM[[ \hbar ]]:B\a \in \CM\{ \hbar \} \}.$$
In the sequel, we shall be concerned with the $\Qt$-algebra but most results
and constructions can be adapted for the algebra $\widehat \Qt$ for which
most proofs are straightforward.\\
That $\Qt$ is a ring follows from general results due to Boutet de
Monvel and Kr{\'e}e (see Proposition \ref{P::BdM}).\\
There is a Borel transform with parameters
that we denote in the same way
$$ B:\widehat \Qt[[ \l ]] \to \CM[[\hbar ,\l, x,y]],\ f \mapsto \sum_{m,n,k \geq 0}
\frac{ \a_{mnk}}{k!}x^my^n\hbar^k,\ \ \a_{mnk} \in \CM[[\l]],$$
where $\widehat \Qt[[ \l ]]$ denotes the algebra of formal power series in
$q,p,\hbar,\l=(\l_1,\dots,\l_k)$ such that the $\l_i$'s are central element of the algebra.
We denote by $\Qt \{ \l \}$ the algebra of elements having a convergent
Borel transform in the $\hbar$ variable.\\
Similarly, we define the ring  $ \Cq \{ \l \}$ as the subring
consisting of formal power series $\sum_{j,k}
\a_{j,k}\l^j \hbar^k$ for which the series $\sum_{j,k}
\frac{\a_{j,k}}{k!}\l^j \hbar^k$ is convergent in a sufficiently small
neighbourhood of the origin. If $k=1$, we often write $t$ or $z$ instead of
$\l$.
\subsection{Statement of the quantum Morse lemma}
Denote by $\widehat \Et_{x,\hat t}(0)$  the ring of germs of formal
microlocal differential operators of order $0$ in the variable
$x$ and $t \in \CM$ which are independent on $t$.
As observed by Pham, the mapping
$$\widehat \Qt \to  \widehat \Et_{x,\hat t}(0),\ (a^\dag,a,\hbar ) \mapsto
(x,(\d_t)^{-1} \d_x,(\d_t)^{-1})$$
is an isomorphism of $\CM$-algebras (\cite{Pham_resurgence}). Therefore, we can apply the results
of algebraic analysis in the $\Qt$-algebra.
In particular, \cite{Boutet_Kree} Lemma 1.2 and Proposition 1.2 imply the following result.
\begin{proposition}[\cite{Boutet_Kree}]
\label{P::BdM}
{The vector space $\Qt \{ \l \}$ is a subalgebra of $\widehat \Qt [[\l ]]$. Moreover for any $u \in \Cq \{\l, z \}$,$u=\sum_{n \geq 0} u_n z^n$,  and any $f \in \Qt\{ \l \}$,
the element $  u \circ f:=\sum_{n \geq 0} u_n f^n$ belongs to the algebra $ \Qt\{ \l \}$.}
\end{proposition}

The proof of the analyticity of perturbative expansions for the
spectrum of a perturbed harmonic oscillator is based on the following theorem.
\begin{theorem}
\label{T::Morse}{For any deformation $f=f_0+tg$ with $g \in \Qt \{ t \}$, there exist an automorphism $\p \in Aut(\Qt\{ t
  \} )$  and a function germ $u \in \Cq \{t,z \}$ such that $u \circ \p(f)=f_0$ provided that the principal symbol of $f_0 \in \Qt$
is a Morse function-germ \footnote{A holomorphic function germ $F:(\CM^n,0) \to \CM$ is called of {\em Morse type}
if  $dF(0)=0$ and $d^2F(0)$ is a non-degenerate quadratic form.}.}
\end{theorem}
The proof of this theorem will be given in Section \ref{S::proof}. 
\begin{remark}{ In the terminology of micro-local analysis, the theorem
 states that any deformation $f$  of an $\hbar$-pseudo differential
operator $f_0$ with a non-degenerate
quadratic part is trivial, i.e., there exists a Fourier integral operator $U_t$
such that
$$U_t f U_t^{-1}=u^{-1}(f_0).  $$
In the limit $\hbar \to 0$, the theorem gives the Vey isochore Morse lemma (\cite{Vey_isochore}); and
if we consider only formal power series in $\hbar$, then the formal variant of the theorem is equivalent to this isochore Morse lemma
(\cite{Colin_Parisse,Helffer_Sjostrand}).}
\end{remark}
By taking a linear interpolation between $f_0 \in \Qt$ and its
quadratic part, we deduce the following corollary.
\begin{corollary}
\label{C::HS}
Under the assumptions of the previous theorem,
there exist an automorphism $\p_0 \in Aut(\Qt )$ 
and a function germ $u_0 \in \Cq \{ z \}$ such that the equality $u_0 \circ
\p_0(f_0)=p^2+q^2$ holds.
\end{corollary}
Corollary \ref{C::HS} was obtained by Helffer and Sj{\"o}strand with the additional assumption that the operator is self-adjoint 
(\cite{Helffer_Sjostrand}, Th\'eor\`eme b1 and Th\'eor\`eme b6).
\section{Differential calculus in the $\Qt$-algebra}
We use the old-fashioned notions and notations of quantum mechanics (\cite{source_book}).
For notational reasons, we consider the algebra $\Qt$ but the results of this section admit straightforward generalisations to
the algebra $\Qt\{ \l \}$.

\subsection{Integration of the Heisenberg equations}

By {\em Heisenberg equations}, we mean a non-autonomous
evolution equation of the type
$\dot F=\frac{i}{\hbar}[ F,H], \ F, H \in \Qt\{ t \}$
where the dot denotes the derivative with respect to $t$.
\begin{proposition}
\label{P::integration}
{For any $H \in \Qt\{ t \}$, there exists a unique operator 
$U \in \Qt \{ t \}$ satisfying the equation $\dot U=HU$ with initial condition $U(t=0,\cdot)=1$.
The
  automorphism $\p \in Aut(\Qt \{ t \})$
$$\p:\Qt \{ t \} \to \Qt\{ t \}, f \mapsto
U(\frac{t}{i\hbar})f U^{-1}(\frac{t}{i\hbar}),\ U \in \Qt\{ t \},$$
integrates the Heisenberg equations of $H \in
\Qt\{ t \}$, that is:
$$\frac{d}{dt}\p(f) =\frac{i}{\hbar}[\p(f),  H ],\ \forall f \in \Qt.$$}
\end{proposition}
\begin{proof}
The proposition states, in particular, that in the expression
$U(\frac{t}{i\hbar})f U^{-1}(\frac{t}{i\hbar})$ all meromorphic terms cancel.\\
First, we show that there exists an element $U \in \Qt\{ t \}$ such that
$\dot U=H U$, $U_{\mid t=0}=1$.\\
Denote by $G \in \Qt\{ t \}$ the element obtained by replacing the
complex coefficients in the series expansion of $H$ in the basis
$((a^\dag)^j a^k t^l \hbar^n)$ by their moduli.
\begin{lemma}
The formal power series $V=\sum v_{jklm}(a^\dag)^ja^kt^l \hbar^m$,
$v_{jklm} \in \RM_{\geq 0}$, obtained by integrating the equation $\dot V=G V$, $V_{\mid t=0}=1$,
is a  majorant series\footnote{A series $\sum a_i z^i$, $a_i \in \CM$,
is a majorant series of another series $\sum b_i z^i$, $b_i \in
\CM$, if the modulus of $a_i$ is at least that of $b_i$ for any multi-index $i$.}
for $U=\sum u_{jklm}(a^\dag)^ja^kt^l\hbar^m$,
i.e., $|u_{jklm}| \leq v_{jklm}$ for all indices.
\end{lemma}
\begin{proof}
In the basis $((a^\dag)^i a^j)$, any expression
$(a^\dag)^k a^l (a^\dag)^m a^n$ can be written as a polynomial with
non-negative coefficients:
$$(a^\dag)^i a^j (a^\dag)^k a^l=\sum c_{mnr}(a^\dag)^m a^n \hbar^r,\ c_{mnr}
\geq 0. $$
Therefore, by inserting the series of $U$ inside the equation $\dot U=HU$, we find
that the coefficients $u=(u_{jklm})$ of $U$ in the basis
$((a^\dag)^j a^kt^l\hbar^m)$ are defined by recurrence relations of
the type $u=uc \a$ where $c$ is a triangular (semi-infinite) matrix
with non negative coefficients and $\a$ is the matrix of $H$.
Replacing the elements $\a$ by their moduli, we
get coefficients $v=(v_{jklm})$ each having a moduli at least equal
to the corresponding coefficient in $u$. This proves the lemma. 
\end{proof}
Now, write $G$ as a series with coefficients in $\CM \{ t\}$ and
denote by $\t G$ the series obtained by replacing each of these
coefficients by their supremum norm in a common sufficiently small
neighbourhood of the origin.
A similar argument to that of the previous lemma shows that the
solution of the equation $\dot {\t V}=\t G \t V$ with $\t V_{\mid t=0}=1$ is a majorant
series for $V$.\\
As $\t G$ is $t$-independent, we have $\t V=e^{t\t G}$, therefore $\t V$ and
consequently $V$ and $U$ are Borel analytic. This proves the assertion.\\
The map
$$\p: \Qt \to \widehat \Qt[[ t,t/\hbar^{-1} ]],\ f \mapsto
U(\frac{t}{i\hbar}) f U^{-1}(\frac{t}{i\hbar})$$
integrates formally the Heisenberg differential equations that is as
formal power series
$$\frac{d}{dt}\p(f)=\frac{i}{\hbar}[\p(f),H].$$
Now, in a sufficiently small neighbourhood of the origin,
the total symbol of $\p(f)$ has the following properties
\begin{enumerate}
\item it is holomorphic in the complement of the hyperplane $\{ \hbar
  =0 \}$,
\item its restriction to the hyperplane $t=0$ is holomorphic (since
the restriction to $t=0$ of  $U$ equals one),
\item its partial derivative with respect to $t$ is holomorphic.
\end{enumerate}
This shows that $\p(f) \in \Qt\{ t \}$ for any $f \in \Qt$ and concludes the proof of the proposition.

\end{proof}
This proposition shows that a change of polarisation induces an
automorphism of the $\Qt$-algebra. For instance, if the series $\sum
\a_{mn}(a^\dag)^ma^n$ lies in $\Qt$ then so do the series $\sum
\a_{mn}a^m(a^\dag)^n$,  $\sum \a_{mn}q^mp^n$ and $\sum
\a_{mn}p^mq^n$.
\subsection{Derivations in $\Qt\{ t \}$}
Following Born, Jordan and Heisenberg (\cite{BHJ}), we define partial
derivatives
$$\d_{q}f=-\frac{i}{\hbar}[ f,p ],\ \d_{p}f=\frac{i}{\hbar}[ f,q ], \
f \in \Qt,\ i^2=-1. $$
We denote by $\int f dq$ (resp.  $\int f dp$) the only function-germ
$F\in \Qt$ such that
\begin{enumerate}
\item $ \d_{q}F=f$ (resp. $\d_{p}F=f $),
\item $F$ is divisible by $q$, i.e., there exists $G \in \Qt$ such that $F=qG$ (resp. $F=p G $).
\end{enumerate}
For instance, if we write $f=\sum_{m,n \geq 0} \a_{mn} q^m p^n$
we get
$$\int f dq=\sum_{m,n \geq 0} \frac{\a_{mn}}{m+1} q^{m+1}
p^n,\ \ \ \d_{q}
f=\sum_{m,n \geq 0} ma_{mn} q^{m-1} p^n.$$

A {\em derivation}  $D:\Qt \{ t \} \to \Qt \{ t \}$  of the algebra
 $\Qt \{ t \}$ is  a $\Cq$-linear mapping satisfying
the Leibniz rule. Due to the non-commutativity of the algebra $\Qt \{ t \}$, the
    space of $\Qt\{ t \}$-derivations is not a $\Qt \{ t \}$-module but only a $\Cq$-module.
\begin{proposition}
\label{P::infinitesimal}
{For any derivation $D$ of the algebra $\Qt \{ t \}$,
there exist function germs $H \in \Qt \{ t \}$ and $\a \in \Cq \{ t \} $
such that $D= \frac{i}{\hbar}[ \cdot, H ]+\a \d_t $.
The function germ $H$ is related to the derivation $D$ by the formula
$$H=\int(Dp )dq -\int(Dq )dp +\frac{i}{\hbar}\int \int [p ,Dq  ]dp  dq .$$}  
\end{proposition}
\begin{remark}{ One can define a
non-commutative de Rham complex in the most obvious manner. This complex
defines a resolution of the ring $\Cq$. The existence of $H$ is an easy
consequence of this fact.}
\end{remark}
\begin{proof}
Define $F=\int (Dq)dp$, then we have the equality $Dq=\d_{p}F=\frac{i}{\hbar}[q,F ]$.
I assert that the function germ $Dp -\frac{i}{\hbar}[p ,F ]$ does not depend on $p
$, that is,  $[ Dp ,q ]=\frac{i}{\hbar}[[p ,F ],q ]$.
As $D$ is a derivation, we have the equalities
$$D[ p ,q  ]=[ Dp ,q  ]+[ p , Dq  ]=0;$$
from which we deduce that $[ Dp ,q ]=[ Dq ,p ]= \frac{i}{\hbar}[[q,F ],p ]$.
Finally, using the Jacobi identity, we get that $[[q,F ],p ]=[[p ,F ],q]$.
This proves the assertion.\\
The assertion implies that $F'=\int  (Dp -\frac{i}{\hbar}[p ,F ])dq $ is a function
of $q$ independent on $p$. We put $H =F +F' $, then
$Dq =\frac{i}{\hbar}[ q ,H]$ and $Dp =\frac{i}{\hbar}[ p ,H]$. Consider the derivation $D'=D-\frac{i}{\hbar}[\cdot,H]$. As one has the equalities $qD'(t)=D'(qt)=D'(t
q)=D'(t)q$ and similarly for $p$,
the function $D'(t)=\a$ belongs to the centre of the algebra
$\Qt \{ t \}$, that is, $\a \in \Cq \{ t \}$.
This shows that the derivation $D'$ is given by $D'=\a \d_t$
and concludes the proof of the proposition.
\end{proof}

\subsection{Infinitesimal action for inverse images}
The action of the ring $\Cq \{z,t \}$ on the space $\Qt\{ t \}$
induces an infinitesimal action that we shall now describe.\\
To the germs of two operators $f,v \in \Qt \{ t \}$, we associate an element $f_v \in
\Qt \{ t,\e \}/(\e^2)$ defined by the rule $f_v(t,q,p)=f(t,q,p)+\e v(t,q,p). $
Here $(\e^2) \subset \Qt \{ t,\e \}$ denotes the ideal generated by $\e^2$.\\
For any element $u \in \Cq \{ t \}$, we write
$$u \circ f_v(t,q,p)=u \circ f(t,q,p)+l(u,f,v) \e $$
with $l(u,f,v) \in \Qt \{ t \}$.\\
We define  the mapping $D_zu(f):\Qt \{ t \} \to \Qt \{ t \},\ v \mapsto l(u,f,v)$. If the operator germ $v$ commutes with $f$ then the equality
$D_z u ( f)\cdot v=(\d_z u \circ f)v$ holds.\\
For instance, for $u(z)=z^n$ we get $D_z u ( f)\cdot v=f^{n-1}v+f^{n-2}vf+\dots+vf^{n-1}$ and if $v$ commutes with $f$ then $D_z u ( f)\cdot v=nf^{n-1}v$.\\
We have the chain rule formula
$$\frac{\d}{\d t} (u \circ f)= (\frac{\d}{\d t} u) \circ f+D_zu(f)\cdot \frac{\d}{\d t}f.$$
If $u$ is invertible for the composition law then the inverse of its derivative satisfies
the equality $D_z u(f) D_zu^{-1}(u \circ f)={\rm Id}$.
\section{Proof of the quantum Morse lemma (Theorem \ref{T::Morse})}
\label{S::proof}
\subsection{Infinitesimal formulation of the quantum Morse lemma} \ \\
As suggested by Thom, for the case of singularity theory for mappings, we start by using the path method (\cite{AVG}).\\
{\em Notations.}{ The $\Cq$-module $M=\Qt\{ t \}/\frac{i}{\hbar}[ f,
    \Qt\{ t \} ]$ has a $\Cq\{ z,t
\}$-module structure induced by that of $\Qt\{ t \}$ :
$$\sum_{n \geq 0}a_n z^n \circ [m]:=[ \sum_{n \geq 0} a_n f^nm],\ a_n \in \Cq \{
 t \},\ m \in \Qt \{ t \},$$
where the brackets $[\cdot ]$ mean that we project the element in
 $M$.}
\begin{lemma}
\label{L::homologique}{The quantum Morse lemma (Theorem \ref{T::Morse})  holds provided that the $\Cq\{
 t\}$-module $M=\Qt\{ t \}/\frac{i}{\hbar}[ f,\Qt\{ t \} ]$ is of finite type.}
\end{lemma}
\begin{proof}
We search for an automorphism $\p \in Aut(\Qt\{ t \})$ with $\p(t)=t$ and a map $u
\in \Cq \{ z,t \}$ such that $u \circ \p(f)=f_0$.
We differentiate this equality with respect to $t$,
we get the equation
\begin{equation}
\label{E::homolo}
\frac{\d u}{\d t} \circ \p(f)+D_z u ( \p(f))\cdot \p(\frac{i}{\hbar}[f, H
]+\frac{\d f}{\d t})=0
\end{equation}
where, according to Proposition \ref{P::infinitesimal}, the operator
$H$ is defined by the equality
$[\cdot,H] +\d_t =\p^{-1}(\frac{d}{dt} \p( \cdot))$. 
Applying the map $D_z u^{-1}(u \circ \p(f))$ to Equation
(\ref{E::homolo}) and then acting by the automorphism $\p^{-1}$, we get an equation of the type
\begin{equation}
\label{E::homologique}
g \circ f+\frac{i}{\hbar}[ f,H ]=\g,\ \g \in \Qt\{
t \},\ g \in \Cq \{ z,t \}.
\end{equation}
The automorphism $\p$ is obtained from 
$H$ by integration of the Heisenberg differential equations (Proposition
\ref{P::integration}). I assert that the map germ $u$ can also be recovered from the map germ $g$.
Indeed, as $g \circ f$ commutes with $f$, the relation
 $$\frac{\d u}{\d t} \circ f=D_z u(f) \cdot (g \circ f)$$
reduces to 
 $$\frac{\d u}{\d t} \circ f=(\frac{\d u}{\d z} g) \circ f.$$
A straightforward variant of the Cauchy-Kovalevska\"\i a theorem with coefficients
in $\Cq$ implies that the initial value problem 
$$
\begin{cases}
 \frac{\d u}{\d t}=\frac{\d u}{\d z} g \\
   u(t=0,z)=z
\end{cases}
$$ can be solved in $\Cq\{ z,t \}$. This fact can also be reduced to the standard Cauchy-Kovalevska\"\i a theorem.
Indeed, the Borel transform $Bu$ of $u$ satisfies the equation
\begin{equation}
\label{E::eq}
\frac{\d Bu}{\d t}=\frac{\d Bu}{\d z} \ast B g,
\end{equation}
with initial condition $Bu(t=0,z)=z$. (We used the relation $B(\a \b)=B\a \ast B\b$ where
$\ast$ denotes the convolution product in the $\hbar$ variable.)\\
Expanding both sides of the differential equation as power series in $\hbar$,
we get that there exists a unique formal power series solution
to this initial value problem; it remains to prove that it is holomorphic.\\
We chose $r \in \RM$, so that,
in the series expansions 
$$Bg=\sum_{n,m \geq 0}g_{n,m}z^nt^m,\ Bu=\sum_{n,m \geq 0}u_{n,m}z^nt^m$$
the coefficients $g_{n,m},u_{n,m} $ are holomorphic in the disk
$D=\{ \hbar \in \CM, |\hbar|<2r \}$ and both functions are holomorphic in some
polydisk $D \times D' \subset \CM^3$.\\
We define the holomorphic function 
$$\t g=\sum_{n,m\geq 0}c_{n,m} z^nt^m,\ c_{n,m}=\sup_{|\hbar| \leq r}|g_{n,m}(\hbar)|.$$
The integro-differential equation (\ref{E::eq}) gives the recursion
$$(m+1)u_{n,m+1}=\sum_{j+j'=n,k+k'=m}(j+1)u_{j+1,k} \ast g_{j',k'}. $$
Therefore an induction on $m$ shows that the solution $v$ of the partial differential equation
 $$\frac{\d v}{\d t}= \frac{\d v}{\d z}\,\t g$$
with initial condition $v(t=0,z)=z$ is a majorant series for $Bu$, that is, the
coefficients of the power series expansions in $z,t$ of the expansion
$v$ majorate that of $Bu$ inside the disk $D$. By the standard Cauchy-Kovalevska\"\i a theorem, the function $v$ is holomorphic in some neighbourhood of the origin
and therefore so is $Bu$. This proves the assertion.\\
This shows that there exist $u,\p$ such that $u \circ \p(f)=f_0$ provided that
there exist $g,H$ satisfying Equation (\ref{E::homologique}).\\
In the notations introduced at the beginning of this subsection,
Equation (\ref{E::homologique}) becomes
 $g \circ [1]=[\g]$.\\
Therefore it can be solved provided that $[1]$ generates
the module $M$.\\
Assume that the module $M$ is of finite type then the Nakayama lemma
implies the equivalences \\
(i) the class of $1$ generates the $\Cq \{z,t \}$-module  $M$,\\
(ii)  the class of $1$ generates the $\CM$-vector space
$V=M/(tM+\hbar M+zM)$.\\

The $\CM$-vector space $V$ is of dimension $1$. Indeed, this vector
space is the fibre at the origin of the Brieskorn lattice of the
symbol $\sigma( f_0)$ of $f_0$
(see e.g. \cite{isochore}). Therefore, the dimension of $V$ equals
the Milnor number of the plane curve singularity $\{ \sigma( f_0)=0
\}$ (\cite{Br3,Malgrange}). The Milnor number of a Morse function is
equal to one; the class of $1$ in $V$ being obviously non-zero, point (ii) is
satisfied and concludes the proof of the lemma.
\end{proof}
\subsection{Finiteness theorem}
\label{SS::sheaves}
By Lemma \ref{L::homologique}, it remains to prove that $M$ is a module of
finite type. We will prove
the following more general theorem (remark that for formal power
series the proof of this theorem is straightforward).
\begin{theorem}
\label{T::finite}{For any germ $f \in \Qt \{ \l \}$,
  $\l=(\l_1,\dots,\l_k)$, such that the
  symbol of $f_0=f_{| \l=0}$ has an isolated critical point at the origin, the space $M=\Qt\{ \l \}/\frac{i}{\hbar}[ f,\Qt\{ \l \}]$ is
  a $\Cq\{ z,\l \}$-module of finite type.}
\end{theorem}
This theorem is a particular case of a finiteness theorem that we shall now formulate.\\
We denote by $\OM_{\CM^{k}|\CM^{k+2}}$ the restriction of the
sheaf of holomorphic functions in $\CM^{k} \times \CM^N$ to the vector subspace $\CM^{k} \times \{ 0 \}$.
The {\em quantum analytic sheaf relative to the projection $(\l,x,y) \mapsto
  \l$}, denoted $\Qt_{\CM^{k+2}/\CM^k}$, is defined by the presheaf  (\cite{Pham_resurgence,Schapira_Polesello}):
$$ U \to \Qt_{\CM^{k+2}/\CM^k}(U)=\{ f \in \widehat \Qt[[\l]] , Bf \in
 \OM_{\CM^{k+2}|\CM^{k+3}}(U) \}$$
where $U$ denotes an open subset.
The sheaf of vector spaces $\Qt_{\CM^{k+2}/\CM^k}$ and $\OM_{\CM^{k+2}|\CM^{k+3}}$
are isomorphic.\\ 
The sheaf $\Bt_{\CM^l}$ is defined on $\CM^l$ by  the presheaf:
$$ U \to \Bt_{\CM^l}(U)=\{ f \in \CM[[\hbar,\a_1,\dots,\a_l]] , Bf \in
 \OM_{\CM^{l}|\CM^{l+1}}(U) \}.$$ 
\begin{definition} Consider a map $F:X \to S, S \subset \CM^l$, satisfying Thom's $a_F$ condition.
A  sheaf $\Ft$ is called {\em  $F$-constructible} if the following condition holds:
 for each point $x \in X$ there exists a neighbourhood $U$ inside the strata of $x$
such that
$$\Ft_{|U} \approx F^{-1}(F_{|U})_* \Ft. $$
\end{definition}
We use the notations introduced sin Subsection \ref{SS::sheaves}.
A complex of coherent $\Qt_{\CM^{k+2}/\CM^k}$-sheaves is called $F$-constructible if its cohomology sheaves are $F$-constructible and if its differential is $F^{-1}\Bt_S$-linear where $\Bt_S $ denotes the restriction of the sheaf
$\Bt_{\CM^l}$ to $S$.
The proof of the following theorem is given in the appendix (see also \cite{finitude}).
\begin{theorem}
\label{T::finitude}
 {Let $F:(\CM^k \times \CM^2,0) \to (\CM^l,0)$ be a holomorphic map germ satisfying the $a_F$-condition.
The cohomology spaces $H^k(K^\cdot) $ associated to a complex of $F$-constructible
$\Qt_{\CM^{k+2}/\CM^l,0}$-coherent modules are $F^{-1}\Bt_{\CM^l,0}$-coherent modules.}
\end{theorem}
\subsection{Proof of Theorem \ref{T::finite}}
\label{SS::proof}
Consider the unfolding of the plane curve singularity associated to the principal symbol of $f$
$$F: (\CM^k \times \CM^2,0) \to   (\CM^{k} \times \CM,0),\ (\l,x,y) \mapsto  (\l,\s( f)(\l,x,y))$$
where $\s$ stands for the principal symbol.
As $F$ defines an isolated complete intersection singularity it admits standard representatives
(sometimes called good or Milnor representatives), which trivially satisfies the Thom $a_F$ condition for any Whitney stratification
which refines the stratification by the rank (see \cite{AVGII,Looijenga}).\\
Let $F:X  \to S,\ (\l,x,y) \to  (\l,\s( f)(\l,x,y))$
be such a representative.\\
We consider the complex of sheaves on $X$:
$$\Kt^\cdot:0 \to \Qt_{X/S} \to \Qt_{X/S} \to 0$$ where the only
non zero boundary map is given by $ g \mapsto \frac{i}{\hbar} [g,f]$. Here $\Qt_{X/S}$ denote the restriction of the sheaf $\Qt_{\CM^{k+2}/\CM^k}$ to $X$.\\
According to Theorem \ref{T::finitude}, it suffices to prove the following lemma.
\begin{lemma}{The sheaf complex $\Kt^\cdot$ is $F$-constructible, i.e, its cohomology sheaves are
locally constant along the fibres of $F:X \to S$.}
\end{lemma}
\begin{proof}
As the fibres of $F$ have at most isolated singular points, it suffices to prove the lemma at regular points of $F$ (any sheaf
restricted to a point is constant).\\
At the level of zero cohomologies, there is nothing to prove, indeed a
coboundary $m \in \Kt^0(X)$ satisfies
$[m,f]=0$ and is therefore constant along the fibres of $F$.\\
Denote by $ \Phi$ be the automorphism of
$\Qt_{\CM^{k+2}/\CM^k}\{ t \}$ obtained by integrating the Heisenberg
equations of $f$. The principal symbol $\p$ of $\Phi$ is the flow of the Hamilton vector field associated to $F$.\\
Now, take a coboundary $m \in \Kt^1(U)$ where $U$ is a sufficiently small open neighbourhood of a regular point of $F$, so that:
\begin{enumerate}
\item it does not contain the origin,
\item the map $\psi:U \to \CM \times S,\ z \mapsto (t,F(z))$ with $\p(t,w)=z$ is biholomorphic onto its image, i.e, $t$ is a local coordinate on the fibres of
the map $F_{|U}$.. 
\end{enumerate}
Define $m_t=\Phi_{t}(m) \in  K^1(\p_t(X))$.
We differentiate $m_t$ with respect to $t$ and use the fact
that $ \Phi_t(f)=f$, we get
$$\frac{d}{dt}(m_t)=  \Phi_t(\frac{i}{\hbar}[ m,f ])= \frac{i}{\hbar}[
\Phi_t(m),f ]. $$
Thus, the cycles $m$ and $m_t$ are cobordant. This shows that
$\Ht^1(\Kt^\cdot)_{\mid U}=F^{-1}(F_{|U})_*\Ht^1(\Kt^\cdot) $ and hence the
complex $\Kt^\cdot$ is $F$-constructible. This proves the lemma and concludes the proof of the theorem.
\end{proof}

\subsection{Complementary result 1: freeness of the deformation module}
\begin{proposition}
\label{P::free}{For any germ $f \in \Qt \{ \l \}$ (resp. $f \in \widehat \Qt [[ \l ]]$),
  $\l=(\l_1,\dots,\l_k)$, such that the
  symbol of $f_0=f_{| \l=0}$ has an isolated critical point at the
  origin, the  $\Cq\{ z,\l \}$-module $M=\Qt\{ \l \}/\frac{i}{\hbar}[
  f,\Qt\{ \l \}]$ (resp. the $\CM[[\hbar, z,\l ]]$-module
$\widehat M=\widehat \Qt[[ \l ]]/\frac{i}{\hbar}[
  f,\widehat \Qt[[ \l ]]\,]$) is free.}
\end{proposition}
\begin{proof}
The proof is a simplified version of that given in \cite{mutau} in
the semi-classical limit. We make the proof in the analytic case, it
differs from the formal case only in notations.\\
We put $\hbar=\l_{k+2}$, $z=\l_{k+1}$ and define the complex $K^\cdot_j$ inductively
by $K^\cdot_{j+1}:=K^\cdot_j/\l_{j+1}K_j^\cdot$ and $K^\cdot_0:=K^\cdot$
(the complex $K^\cdot$ was defined in Subsection \ref{SS::proof}). The
multiplication by $\l_{j+1}$ induces an exact sequence of complexes
$$0 \to K^\cdot_j \to K^\cdot_j \to K^\cdot_{j+1} \to 0$$ which induces in
turn a long exact sequence in cohomology. There are canonical
isomorphisms
$$H^0(K^\cdot_j) \approx  \Cq\{\l_{j+1},\dots,\l_{k+1}
\},\ H^0(K^\cdot_{k+1}) \approx  \Cq,\ \ H^0(K^\cdot_{k+2}) \approx  \CM$$
for $ j=0,\dots,k$.
Therefore the exact sequences split and we get short exact sequences
$$ 0 \to H^1(K^\cdot_j) \to H^1(K^\cdot_j) \to H^1(K^\cdot_{j+1}) \to 0 $$
which shows that $(\l_1,\dots,\l_{k+2})$ is a regular sequence of maximal length,
therefore the finite type module $M=H^1(K^\cdot)$ has depth $k+2$; consequently the
Auslander-Buchsbaum formula implies that $M$ is a free module (see
e.g. \cite{Eisenbud}).
\end{proof}
\subsection{Complementary result 2: the quantum versal deformation theorem}
The finiteness of the deformation module (Theorem \ref{T::finite}) implies
the versal deformation theorem in the algebra $\Qt\{ \l \}$. The proof
is similar to the one we gave in the isochore case(\cite{isochore}).\\
We recall some standard definitions adapted to our setting.\\
An element $F \in \Qt\{ \l \}$ is called a {\em deformation} of $f=F(0,\cdot)
\in \Qt$. A deformation $G \in \Qt\{\mu\}$ of $f$ is called induced from $F$ is
there exist homomorphisms of algebras $\p:\Qt\{ \l \} \to \Qt\{ \mu
\}$, $u \in \Cq\{ \mu \}$ such that $u \circ G= \p(F)$.\\
A deformation of $f \in \Qt$ is called {\em versal} if any other
deformation of $f$ can be induced from it.
\begin{theorem}[compare \cite{Colin}, Theorems 6,7,8,9 and \cite{Pham}]
\label{T::versal}
{A deformation $F$ of an element $f \in \Qt$ is versal 
provided that the classes of the
$\d_{\l_j}\s(F)_{\mid \l=0}$'s and of $1$ generate the $\CM$-vector
space $\CM\{ x,y \}/(\{ \CM\{ x,y\},\s(f) \} +  \CM\{ x,y
\}\s(f))$.}
\end{theorem}
\begin{remark}{The converse statement of the above theorem holds trivially.}
\end{remark}
\begin{example}The deformation $F=p^2+q^{k+1}+\sum_{j=1}^{k-1} \l_j
  q^j$ is versal. Indeed, here $\s(f)=y^2+x^{k+1}$ and the $\CM$-vector space
$\CM\{ x,y \}/(\{ \CM\{ x,y\},\s(f) \} +  \CM\{ x,y
\}\s(f))$ can be identified with the Milnor algebra $\CM\{ x,y \}/(y,x^k) $
of $f$ which is generated by the classes of $1,x,\dots,x^{k-1}$
(see \cite{isochore}, Example 2 for details).
\end{example}
\begin{proof}
We use a standard method introduced by Martinet in the context of singularity theory for differentiable mappings \cite{Mar}
(see also \cite{isochore}).\\
Let $G$ be an arbitrary deformation of $f$ depending on the parameters
$\mu_1,\dots,\mu_l$.\\
Define the deformation
$\Phi=F+G-f$ and let $\Phi_j$ be the restriction of $\Phi$ to
$\mu_1=\dots=\mu_j=0$ with $\Phi_0=\Phi$.\\
{\em Assertion.} The deformation $\Phi_{j-1}$ is induced by the
deformation $\Phi_j$.\\
We put $t=\mu_j$, $\a=(\l_1,\dots,\l_k,\mu_1,\dots,\mu_{j})$ and differentiate with
respect to $t$ the equation $u_t \circ
\p_t(\Phi_{j-1})=\Phi_j$. Proceeding like in the proof of the quantum
Morse lemma, we get the equation
\begin{equation}
\label{E::Colin}
g \circ \Phi_{j-1}+\frac{i}{\hbar}[\Phi_{j-1},H ]+
\sum_{l=1}^{k+j-1}a_l \d_{\a_l}\Phi_{j-1}=\g.
\end{equation}
with $\ \g \in \Qt\{\a \},\ g \in \Cq \{ z,\a \},\ a_l \in \Cq \{ \a \}$.
This equation can be solved provided that $[1]$ and the $[\d_{\a_l}\Phi_{l-1}]$'s generate
the $\Cq\{ \a,z \}$-module
$M=\Qt\{\a \}/\frac{i}{\hbar}[ \Phi_{j-1}, \Qt\{ \a \} ]$.\\
Theorem \ref{T::finite} implies that
the module $M$ is of finite type, therefore the Nakayama lemma
implies the equivalences \\
(i) the classes of $1$ and of the $\d_{\a_l}\Phi_{j-1}$'s generate the $\Cq \{z,\a \}$-module  $M$,\\
(ii)  the classes of $1$  and of the $\d_{\a_l}\Phi_{j-1}$'s generate the $\CM$-vector space
$V=M/\M M$ where $\M$ is the maximal ideal of the local ring $\Cq \{z,\a \}$.\\
The assumption on $F$ implies the last statement.
This proves the assertion.\\
Applying successively the assertion from
$j=0$ to $j=l$, we get that $ \Phi_0=F+G-f$ is induced by $\Phi_l=F$.
This concludes the proof of the theorem.
\end{proof}
\subsection{Complementary result 3: solution to a conjecture due to
  Colin de Verdi{\`e}re}
In \cite{Colin}, Colin de Verdi{\`e}re conjectured the following result.
\begin{theorem}
\label{T::unicite}
Let $F \in \Qt \{ \l \}$ (resp. $F \in \widehat \Qt [[ \l ]]$)
be a miniversal deformation\footnote{A versal deformation depending on
a minimal number of parameter is called {\em miniversal}} of an
operator $f \in \Qt$ (resp. $f \in \widehat \Qt$).
Let $G$ be another deformation of $f$, so that $G$ is induced from $F$, that is,
$u \circ G=\p(F)$, then the function germs $\p(\l_j) \in \Cq\{ \mu \}$ and $u
\in \Cq \{ \mu,z \}$
(resp. $\p(\l_j) \in \CM[[\hbar,\mu ]]$ and $u \in \CM[[\hbar,\mu,z
]]$) are uniquely determined by the choices of $F$ and $G$.
\end{theorem} 
\begin{proof}
We make the proof for the analytic case, the case of formal power
series is similar (and in fact simpler since the finiteness of the
deformation module is in this case obvious).\\
We use the same notations as in the proof of Theorem
\ref{T::versal}.\\
Equation  (\ref{E::Colin}) can be written as
\begin{equation}
\label{E::Colin2}
g \circ [1]+
\sum_{l=1}^{k+j-1}a_l [\d_{\a_l}\Phi_{j-1}]=[\g]
\end{equation}
where the bracket denotes the class in  the module
$M=\Qt\{\a \}/\frac{i}{\hbar}[ \Phi_{j-1}, \Qt\{ \a \} ]$.\\
Since $F$ is miniversal and the module $M$ is free of
finite type, the classes $[\d_{\a_l} \Phi_{j-1}]$ and $[1]$ freely generate the
module $M$ for $l \in \{ 1,\dots,k \}$.
Therefore the solution of Equation (\ref{E::Colin}) with $a_l=0$ for $l > k$
is unique.\\
This shows that the functions germs $\p_t^{-1}(\l_k)$ obtained after
integrating the coefficients $a_l$ are uniquely determined.
By a finite induction on $j \in \{0,\dots,k \}$, we get that the
function germs $\p(\l_k)$ are uniquely determined by $F$ and $G$.
This proves the theorem.
\end{proof}
\section{Analyticity of the perturbative expansions}

We come back to the convergence of the perturbative expansions, our
aim is to associate to each element in $\Qt$ a spectrum which is invariant under automorphism connected to the identity
and which preserve analyticity properties.
\subsection{The operator representation.}
\label{SS::representation}
Denote by $\Qt a$ the left ideal generated by $a$ and put $\Ht=\Qt/\Qt
a$. The map
$\Ht \to \Cq \{ z \} $ sending the class of $a^\dag$ to $z$ is an
isomorphism of $\Cq$-modules.\\
The left multiplication by $H \in \Ht$ induces a commutative diagram
defining the homomorphism $\rho(H) \in \Hom_{\Cq}(\Ht,\Ht)$:
$$\xymatrix{\Qt \ar[r]^-{H.} \ar[d] & \Qt \ar[d] \\
             \Ht \ar[r]^-{\rho(H)} & \Ht}$$
Here the vertical arrows stand for the canonical projections.\\
Thus, we have a homomorphism of $\Cq$-modules
$$\rho:\Qt \to \Hom_{\Cq}(\Ht,\Ht)$$
representing the elements in $\Qt$ as $\Cq $-linear operators in
$\Ht$.
Via the isomorphism $\Ht \approx \Cq \{ z \} $, the operators associated to $a^\dag$ and $a$
are mapped to the multiplication by $z$ and to $\hbar \d_z$.
\begin{proposition}
\label{P::monomorphism}
The  homomorphisms of $\Cq$-rings $\rho:\Qt \to \Hom_{\Cq}(\Ht,\Ht)$,
is a ring monomorphisms.
\end{proposition}
\begin{proof}
The kernel $I$ of the homomorphism $\rho$ is a left-ideal
invariant under right multiplication by $a$ and $a^\dag$.\\
Define the map $v:I \to \ZM_{\geq 0}$ sending $H \in I$ to the smallest
$k \in \ZM_{\geq 0}$ for which there exists $j$ such that the coefficient of
$(a^\dag)^ja^k$ in the expansion $H=\sum_{jk} \a_{jk}(a^\dag)^j a^k$ is non zero.\\
Assume that $I \neq 0$, then there exists at least one non-zero
element $H$ for which $v$ is minimal.\\
We have necessarily $v(H)=0$ otherwise $v([H,a^\dag])$ would be smaller than
$v(H)$. Evaluating $H_k=[a^k,H]$ with $H=\sum_{j} \a_{j}(a^\dag)^j$ on the class of $1 \in\Qt$, we get that 
$\rho(H_k) \bar 1 =\hbar^k \a_{k}\bar 1=0 $. This contradicts
the fact that $H \neq 0$. 
\end{proof}

\subsection{The spectrum of an operator}
In the sequel, we identify $H \in \Qt$ with its image under the homomorphism $\rho$.
The set of eigenvalues of an operator $H \in \Qt$ under the $\rho$-representation, denoted by $\Sp(H)$, is
called the {\em analytic spectrum}. There is also a representation
$$\hat \rho:\hat \Qt \to \Hom_{\Cq}(\widehat \Ht,\widehat \Ht)$$
with $\widehat \Ht=\widehat \Qt/\widehat \Qt a$,  the resulting {\em formal spectrum}  of an operator $H \in \widehat \Qt$ is denoted by
$\widehat \Sp(H)$.
\begin{proposition}The analytic spectrum and the formal spectrum of the harmonic oscillator $H=a^\dag a$
coincide and are both equal to
$\hbar \ZM_{\geq 0}$.
\end{proposition}
\begin{proof}
We have $H (a^\dag)^n=n\hbar (a^\dag)^n \ (mod \ \Qt a)$ therefore the
projection of the $(a^\dag)^n$'s in $\Ht$ are the eigenvectors of $\rho(H)$.
\end{proof}

\subsection{Borel analyticity of perturbative expansions}
We now consider the case with parameters.\\
In the space $\Ht\{ t \}:=\Qt\{ t \}/\Qt\{ t \} a$ we cannot take the
definition of the previous subsection for the eigenvectors, since already in example of the harmonic oscillator, perturbative expansions of eigenvectors are, in general, neither holomorphic nor meromorphic. Therefore, we shall say that $\psi \in \Ht\{ t \}$ is an {\em eigenvector} of $f \in \Qt\{ t \}$ if there exists
$E \in \Cq \{ t \}$ such that
$$f(\hbar t,\cdot) \psi(\hbar t,\cdot)=E(t,\cdot) \psi(\hbar t,\cdot). $$ 
Consequently, if $\psi(t,\cdot)$ is an eigenvector then  $\psi(t/\hbar,\cdot)$ is the corresponding perturbative expansion.
Similar consideration apply in the formal case.\\
Let us denote by $Id \in Aut (\Qt\{ t \})$ the identity mapping.
Proposition \ref{P::integration} implies the following result.
\begin{proposition}
\label{P::invariant}
For any automorphism $\p $ of $\Qt \{ t \}, \p(t=0,\cdot)=Id$, any mapping $\psi \in \Cq\{t,z \}$, and any
germ $f \in \Qt\{t \}$, we have
$$\Sp(\p(f))=\Sp(f),\ \Sp(\psi(f))=\psi(\Sp(f))$$
and similarly for the formal spectrum.
\end{proposition}

\begin{theorem}
\label{T::Heisenberg}{If $f \in \Qt\{ t \}$ is a perturbation of a harmonic oscillator, then its formal and analytic spectrum are equal
$$\widehat Sp(f)=Sp(f),$$
that is, the perturbative expansions of the spectrum are Borel analytic.}
\end{theorem}
\begin{proof}
The quantum Morse lemma (Theorem \ref{T::Morse}) asserts that there exist an automorphism
$\p \in Aut(\Qt \{ t \})$ and an element $u \in \Cq \{ t,z \}$ such
that the equality $f=u \circ \p(f_0)$ holds with $f_0=f(t=0,\cdot)$.
Proposition \ref{P::invariant} implies that
 the eigenvalues of $f$ are the
images of the eigenvalues of $f_0$ under the map-germ $u$. These expansions are unique
(Theorem \ref{T::unicite}) therefore $\widehat Sp(f)=Sp(f)=u(Sp(f_0))$.
In particular, the Borel transform of the perturbative expansion for the spectrum
are holomorphic function germs. This proves the theorem.
\end{proof}
\begin{remark} Via the isomorphism  $\Ht \approx \Cq \{ z \}$, the operator $H$ might be identified
with the operator $\hbar z \d_z$.
For any $\l$, the function $z^{\l/\hbar}$ lies in the kernel of the
operator $\hbar z \d_z-\l$ but only for $\l \in \hbar \ZM_{\geq 0}$
is this solution an unbranched holomorphic function germ.
\end{remark}
\begin{remark} The module $\Ht$ is an inductive limit of $C^*$-Hilbert modules over the ring $\Cq$.
Indeed, the ring $\Cq$ admits the involution
$$\tau:\Cq \to \Cq,\ \sum_n \a_n \hbar^n \mapsto \sum_n \bar \a_n \hbar^n$$
which fixed points form a subring $\RM_{\hbar}$
ordered by the condition $f > g$ if and only if the first non vanishing coefficient
in the series of $f-g$ is positive.\\
Consider the ``restriction to zero'' mapping
$$ \pi:\Qt \to \Cq,\ \ \sum_{m,n \geq 0} \a_{mn} (a^\dag)^m a^n \mapsto \a_{0,0}.$$
We define hermitian conjugation in $\Qt$ by 
$$\dag:\Qt \to \Qt, \sum_{m,n \geq 0} \a_{mn} (a^\dag)^m a^n
\mapsto  \sum_{m,n \geq 0} \bar \a_{mn} (a^\dag)^n a^m;  $$
and a pairing $P:\Qt \times \Qt \to \Cq,\ (f,g) \mapsto \pi(f^\dag g)$.\\
The inner product $\< \cdot | \cdot \> $ is defined by the commutative diagram
$$\xymatrix{\Qt \times \Qt \ar[rd]^-P \ar[d] \\
            \Ht \times \Ht \ar[r]_-{\< \cdot | \cdot \> } & \Cq \\}$$
where the vertical arrow denotes the canonical projection.
Moreover, an easy computation shows that {\em the $\hbar$-trace series}
$Tr_\hbar(f ):=\sum_{n \geq 0} \< n | f | n \> $
of any function-germ $f  \in \Qt$ is Borel analytic. Therefore, not only the spectrum is well-defined
in local analytic terms but also a Hamiltonian one dimensional quantum field theory.
\end{remark}
\newpage
\appendix

\section{Proof of the finiteness theorem}
We assume some familiarity with topological tensor products as
exposed in \cite{Grothendieck_these,Verdier} and with the Whitney-Thom theory
of stratified spaces and morphisms as exposed for instance in \cite{Gibson_Looijenga}. An elementary exposition of this appendix reviewing both theories
is given in \cite{finitude}.
\subsection{Construction of the contraction}
We denote by $B_r$ the ball of radius $r$ centred at the origin in
$\CM^{k+2}$.
Let $F:X \to S$ be a standard representative of $F$. Here $X$ is the
intersection of a Stein open neighbourhood $Y$ of the origin with a small closed ball $B_\e$.
The aim of this subsection is to prove the following proposition.
\begin{proposition}
\label{P::contraction}
For any $\e'<\e$, the restriction mapping $r:\Kt^\cdot(X) \to \Kt^\cdot(X')$ is a quasi-isomorphism
with $X'=X \cap  B_{\e'}$.
\end{proposition}
 It is obviously sufficient to prove
the proposition for $\e'$ sufficiently close to $\e$.
\begin{proof}
As the map $F$ satisfies Thom's $a_F$ condition, there exists a stratified vector field $\theta$ in $Y \subset \CM^n$ tangent to the fibres of $F$
every where transversal to $Y \cap B_\e$.
Denote by $\p:]-\dt,\dt[ \times B_\e \to X$ the flow of the vector field $\theta$ where $\dt$ is small enough so that it
induces a map which is a homeomorphism onto a neighbourhood of the boundary of $B_\e$ as stratified sets:
$$]-\dt,\dt[ \times \d B_{\e} \to B_\e,\ (t,x) \mapsto \p_t(x).$$ 
 We chose $\e'$ sufficiently close to $\e$ so that the boundary of $X'$ lies on the image of this mapping.
Chose an acyclic covering $U=(U_i)$ of $X$, its image $U'=(U_i')$, $U_i'=U_i \cap B_{\e'}$, is an acyclic covering of $X'$.\\
Consider the spectral sequences $E_0^{p,q}(X)=\Ct^p(U,\Kt^q)$,
$E_0^{p,q}(X')=\Ct^p(U',\Kt^q)$ for the hypercohomology
of the complex $\Kt^\cdot$. Here, as usual, $\Ct^\cdot(\cdot)$ stands for the
\v{C}ech resolution.\\
 The map $\p$ induces a homeomorphism between each strata in $U_i$ and the
corresponding stratum in $U_i'$ for each $i$.
As the complex of sheaves is $F$-constructible and $F(U_i)=F(U_i')$
we have group isomorphisms
$$\Ht^q(\Kt^\cdot)(U_i)  \approx F_*\Ht^q(\Kt^\cdot))(F(U_i))\approx \Ht^q(\Kt^\cdot)(U_i')$$ on each small open subset $U_i$.
Therefore, the restriction mapping induces an isomorphism between the first sheets of the hypercohomology
spectral sequences:
$$E_1^{p,q}(X)=\Ct^p(U,\Ht^q(\Kt^\cdot)) \approx \Ct^p(U',\Ht^q(\Kt^\cdot))=E_1^{p,q}(X').$$
This shows that the hypercohomology spaces
$\HM^\cdot(X,\Kt^\cdot)$ and $\HM^\cdot(X',\Kt^\cdot)$ are isomorphic.\\
As $X$ is Stein, by Cartan's theorem B, for any $p \geq 0$, we have the isomorphisms
$$\HM^p(X,\Kt^\cdot) \approx H^p(\Kt^\cdot(X)),\ \HM^p(X',\Kt^\cdot) \approx
H^p(\Kt^\cdot(X')),$$
therefore the restriction mapping is a quasi-isomorphism. This proves the proposition.
\end{proof}
\subsection{Proof of Theorem \ref{T::finitude}}
 \begin{theorem}[\cite{Houzel}]
\label{T::Houzel}Let $A$ be a multiplicatively convex, complete bornological algebra and let $(M_i^\cdot)$ be a sequence
of complexes of complete bornological $A$ modules; let be given for $1 \leq i \leq r$, a homomorphism of complexes
$u_i:M_{i-1}^\cdot \to M_i^\cdot$ ($A$-linear and bounded), and relative integers $a,b \in \ZM$, with $a \leq b$.\\
We make the following assumptions
\begin{enumerate}
\item for all $i$, $M_i^n$ satisfies the homomorphism property for $n \geq a$ and is zero for $n \geq b$,
\item for $1 \leq i \leq r$, $u_i$ is a quasi-isomorphism (see \cite{Illusie}) and is $A$-nuclear in degree $\geq a$,
\item $r \geq b-a+1$.
\end{enumerate}
Then the complexes $M_i^\cdot$ are $a$-pseudo-coherent (see \cite{Illusie}).
\end{theorem}
We take $r=3$, so that $i=0,1,2$, and $a=0$, $b=2$, so that $M_2$ is zero.\\
As the restriction mapping $r:\Kt^\cdot(X) \to \Kt^\cdot(X')$ is a
$\Bt(S)$-subnuclear quasi-isomorphism, the Houzel theorem applies.
This shows that the cohomology spaces of the modules $\Kt^\cdot(X)$ are $\Bt_{S}(S)$-coherent (see \cite{finitude} for details).
Denote by $\Lt^\cdot$ a complex of free coherent $\Bt_{S}$-sheaves so
that $\Lt^\cdot(S)$ is quasi-isomorphic to $\Kt^\cdot(X)$.
\begin{lemma} The sheaf complexes $\Lt^\cdot$, $ f_* \Kt^\cdot_{|X}$ are quasi-isomorphic
\end{lemma}
\begin{proof}
A mapping $u:M^\cdot \to L^\cdot$ of complexes induces a quasi-isomorphism between
two complexes if and only if its mapping cylinder $C^\cdot(u)$ is exact.\\
We apply this fact to the mapping cylinder of the quasi-isomorphism 
$$u:\Lt^\cdot(S) \to \Kt^\cdot(X).$$ 
As the vector space $\Bt_S(P)$ is nuclear for any polydisk $P \subset S$,
the functor $\hat \otimes\Bt_S(P)$ is exact (\cite{Verdier}). Therefore,
the complex $C^\cdot(u) \hat \otimes \Bt_{S}(P)$ is also exact.\\
The complex
 $C^\cdot(u) \hat \otimes  \Bt_{S}(P)$ is the mapping cylinder of the mapping
$u':\Lt^\cdot(P) \to \Kt^\cdot(X \cap f^{-1}(P))$.
Therefore, the complexes of sheaves $\Lt^\cdot$ and $ f_* \Kt^\cdot_{|X}$ are
quasi-isomorphic. This proves the lemma.
\end{proof}
I assert that the complex $K^\cdot=\Kt^\cdot_0$ is quasi-isomorphic to the stalk of the complex $\Lt^\cdot$ at the origin.\\
Let $(B_{\e_n})$ be a fundamental sequence of neighbourhoods of the origin in $\CM^n$,
so that their intersection with the special fibre of $F$
is transverse. As the map $F$ satisfies the $a_F$-condition, we may find a fundamental sequence $(S_n)$ of neighbourhoods of the origin
in $\CM^k$ so that the fibres of $F$ intersect $B_{\e_n}$ transversally above $S_n$.\\
Put $X_n=f^{-1}(S_n)$, we have the isomorphism
$$\Lt_{|S_n}^\cdot \approx F_* \Kt^\cdot_{|X_n} \approx  F_* \Kt^\cdot_{|X_n \cap B_{\e_n}}. $$
The first isomorphism is a consequence of the previous lemma and the second one follows from the fact that the
contraction is a quasi-isomorphism (Proposition \ref{P::contraction}).\\
In the limit $n \to \infty$, we get that the complex $K^\cdot=\Kt^\cdot_0$ is quasi-isomorphic to the complex $\Lt^\cdot_0$.
This concludes the proof of Theorem \ref{T::finitude}.\\

{\em Acknowledgements.}
The author acknowledge F. Aicardi, L. Boutet de Monvel, Y. Colin de Verdi{\`e}re,
O. Maspfuhl, F. Pham, A. Voros, S. V. Ngoc for stimulating discussions.
The author thanks the referee for his detailed comments. 
Special thanks to J. Zinn-Justin for his teaching, to D. van
Straten for important suggestions during the writing of
this work and to P. Schapira for useful comments.
Thanks also to B. Malgrange for pointing out several omissions in the original version of the text.
\bibliographystyle{amsplain}
\bibliography{master}
\end{document}